\documentstyle[12pt,epsfig]{article}
\newcommand{\PR}{{\it Phys.~Rev.~}}

\newcommand{\NP}{{\it Nucl.~Phys.~}}

\newcommand{\PL}{{\it Phys.~Lett.~}}

\newcommand{\ZP}{{\it Zeit.~Phys.~}}

\newcommand{\vol}[1]{{\bf #1}}\newcommand{\vyp}[3]{\vol{#1} (#2) #3}
\newcommand{\MS}{\hbox{$\overline{\rm MS}$}}
\newcommand{\be}{\begin{equation}}
\newcommand{\ee}{\end{equation}}
\newcommand{\bea}{\begin{eqnarray}}
\newcommand{\eea}{\end{eqnarray}}

\begin{document}
\sloppy
\begin{titlepage}
\samepage{
\setcounter{page}{0}
\rightline{CERN-TH/95-184}
\rightline{hep-ph/9507211}
\vspace{.2in}
\begin{center}
{\bf SCHEME DEPENDENCE AT SMALL $x$\\}
\vspace{.3in}
{Stefano Forte\footnote{
On leave from INFN, Sezione di Torino, Turin, Italy
}and Richard D. Ball\footnote{On leave from a Royal Society University
Research Fellowship}\\}
\vspace{.15in}
{\it Theory Division,
 CERN\\ CH-1211 Gen\`eve 23, Switzerland\\}
\end{center}
\vspace{.25in}
\begin{abstract}
We discuss the evolution of $F_2^p$ at small $x$, emphasizing
the uncertainties related to expansion, fitting,
renormalization and factorization scheme dependence.
We find that
perturbative extrapolation from the measured region down to
smaller $x$ and lower $Q^2$
may become strongly scheme dependent.

\end{abstract}
\vspace{.2in}
\begin{center}
 Presented at Workshop on {\bf Deep Inelastic Scattering and QCD},\\
Paris, April 1995\\
\vspace{.1in}
{To be published in the proceedings}
\end{center}
\vfill
\leftline{CERN-TH/95-184}
\leftline{June 1995}
}
\end{titlepage}
\setcounter{footnote}{0}
It is now well established\cite{reviews} that the behaviour of $F_2$
in the region of small $x$ and large $Q^2$ accessed by the
HERA experiments\cite{data} provides a confirmation of the
double scaling behaviour\cite{DAS} predicted asymptotically in perturbative
QCD\cite{DGPTWZ}. As the experimental accuracy improves, it is now possible to
test the theory beyond this simple leading order prediction, by comparing
the data to a full next-to-leading order (NLO) determination of the $x$
and $Q^2$ dependence of $F_2$. This, however, requires a
study  of the
renormalization and factorization scheme dependence
which characterizes perturbative computations, and which
at small $x$ become particularly significant, due to the  growth of
anomalous dimensions and coefficient functions. Moreover,
the presence in the problem of two large scales  ($Q^2$ and
$s=Q^2(1-x)/x$)
requires the choice of an expansion
scheme which  sums up all the appropriate  leading (and subleading)
logarithms. Here we assess the size of these ambiguities and in particular
discuss how they affect the computation of $F_2$ and, conversely,
the extraction from $F_2$ of information on the form of parton distributions
at small $x$.

Determining the evolution of structure functions
by solution of the renormalization group equations in leading
order corresponds to summing all logs of the form $\alpha_s^p (\log Q^2)^q
(\log{1\over x})^r$ with $p=q$ and $0\le r \le p$; double scaling is
a consequence of the dominance at small $x$ of the contributions
with $r=p=q$, i.e., such that the two large logs are treated symmetrically.
It is in fact possible\cite{summing}
to reorganize the perturbative expansion in such a way
that the full LO contribution to anomalous dimensions treats the two logs
symmetrically, i.e. such that in LO each power of $\alpha_s$
is accompanied by either of the two logs (that is, such that
 $1\le q\le p$, $0\le r\le p$, $1\le p \le q+r$). This expansion scheme
(the double leading scheme) can then be extended to NLO and beyond.
Within any given scheme the structure functions are expressed as power
series in $\alpha_s$, even though solving the renormalization group equations
sums contributions involving large logarithms to all orders in $\alpha_s$.
%

Consistent solution of the evolution
equations in any specified expansion scheme and to a given order
is then (at least in principle) always possible.
In practice, the anomalous dimensions are known through their Laurent
expansion in $N$. All the LO coefficients of this expansion are
known for the $2\times 2$ matrix of singlet anomalous dimensions\cite{fact},
but the NLO coefficients of the singular terms in $N$ are
only known for $\gamma_N^{qg}$ and $\gamma_N^{qq}$\cite{cathaut};
the corresponding coefficients in $\gamma_N^{gg}$ and $\gamma_N^{gq}$
can however be fixed by requiring momentum conservation\cite{alphas}.
We will henceforth consider
NLO computations in the double-leading scheme.\footnote{Notice that this is
not quite the same as the approach of ref.~\cite{EHW},
where the higher order singularities are simply added to the one and
two loop anomalous dimensions: in the double-leading expansion all
the NLO terms may be  treated consistently,
by linearizing them in order to avoid spurious sub-subleading terms.}
More specifically perturbative evolution is performed in the usual
loop expansion scheme down to a certain $x_0$, and then the
double-leading scheme used below it.
The value of the parameter $x_0$ can only be determined by
comparison to experiment. Here we will consider two extreme
double-leading NLO scenarios,
namely $x_0$ smaller than any value of $x$ covered by the HERA data,
(i.e., in practice, $x_0=0$ or ordinary two-loop evolution), and $x_0=0.1$.

Once $x_0$ is fixed, renormalization and factorization schemes still have
to be specified in order to perform NLO computations. Without loss of
generality the renormalization scheme will be chosen to be \MS: other
renormalization schemes then correspond simply to a change of
renormalization scale. The choice of factorization scheme is more complex.
Firstly, we have a choice between schemes in
which to all orders $F_2$ is directly proportional to the quark distribution
(parton schemes, such as the DIS scheme), and schemes where (starting
at NLO) $F_2$ receives a gluon contribution (such as the \MS\ scheme).
Furthermore, in parton schemes the coefficients of NLO singular
contributions to the singlet anomalous dimensions also depend on the
choice of factorization scheme: when computed in the DIS scheme\cite{cathaut}
they contain a process-independent singularity in the quark
sector, which is removed if off-shell factorization is used instead
(Q$_0$DIS scheme)\cite{cia}. It is
even possible to set the NLO singularities in the quark sector
to zero, thereby factorizing the entire singularity into the starting
distribution (SDIS scheme)\cite{sdis}.\footnote{It turns out
that in the HERA region the results obtained in the
SDIS scheme are essentially identical to those found by ordinary two loop
evolution\cite{alphas}.} Corresponding \MS\ schemes may be
constructed by insisting that the anomalous dimensions be the same as those
in the standard \MS\ scheme \cite{cathaut}, but with the coefficients
being adjusted accordingly (so that in particular in Q$_0$\MS\ scheme
the process - independent singularity is removed from the
coefficient function).

\begin{table}[t]\begin{center}
\begin{tabular}{|c|c|c|c|c|}
\hline
       &         norm.(\%)      &$\lambda_q$      & $\lambda_g$     & $\chi^2$
\\ \hline\hline
    & $96\quad 103$ & $-0.23\pm 0.05$ & $0.10\pm 0.07$  & $57.3$*\\
  a)  & $97\quad 103$ & $-0.24\pm 0.05$ & $0.12\pm 0.08$  & $57.6$\\
\cline{2-5}
     & $94\quad 101$ & $-0.24\pm 0.09$ & $-0.52\pm 0.23$ & $59.2$\\
       & $95\quad 101$ & $-0.25\pm 0.10$ & $-0.49\pm 0.26$ & $59.0$*\\
\hline
     & $96\quad 102$ & $-0.25\pm 0.02$ & $0.03\pm 0.16$  & $64.5$*\\
  b) & $97\quad 104$ & $-0.25\pm 0.02$ & $-0.08\pm 0.01$ & $58.1$\\
\cline{2-5}
   & $97\quad 104$ & $-0.12\pm 0.02$ & $-0.01\pm 0.20$ & $62.5$\\
    & $97\quad 103$ & $-0.13\pm 0.07$ & $-0.36\pm 0.24$ & $57.9$*\\
\hline
        & $96\quad 102$ & $-0.26\pm 0.02$ & $0.12\pm 0.17$  & $72.6$*\\
 c)  & $98\quad 106$ & $-0.24\pm 0.02$ & $-0.17\pm 0.09$ & $65.0$\\
\cline{2-5}
       & $97\quad 103$ & $0.10\pm 0.06$  & $0.01\pm 0.37$  & $73.1$\\
       & $95\quad 101$ & $-0.03\pm 0.03$ & $-0.75\pm 0.05$ & $62.3$*\\
 \hline\hline
        & $100\quad 106$ & $-0.13\pm 0.04$ & $0.18\pm 0.01$  & $63.4$\\
  d)   & $94\quad 100$ & $-0.22\pm 0.06$ & $-0.10\pm 0.10$ & $57.9$\\
\cline{2-5}
       & $100\quad 107$ & $-0.16\pm 0.11$  & $0.07\pm 0.25$  & $63.9$\\
      & $89\quad 95$ & $-0.28\pm 0.05$ & $-0.70\pm 0.12$ & $73.9$\\
\hline\hline
\end{tabular}
\caption[]{\label{tabella}
Fitted parameters for: a) $x_0=0$; b) $x_0=0.1$, $Q_0$ factorization;
c) $x_0=0.1$, standard factorization. In each case the four entries
correspond respectively to
DIS distributions (DIS  and \MS\ evolution); \MS\ distributions
(DIS  and \MS\ evolution). The two entries d) show the effect on the first
and fourth
entry of the table of varying the renormalization scale by a factor of two
either side.
}
\vskip-0.25truecm
\end{center}
\end{table}
Finally, there is still an ambiguity in the definition of the initial parton
distributions (a `fitting scheme' ambiguity), related to the fact that
these  can be fitted in a parton scheme or in an \MS\ scheme  regardless
of which scheme is chosen to evolve.
Besides providing information on the dependence of the results
for $F_2$ on the specific choice of parton parametrization,
varying the fitting scheme demonstrates the implicit scheme dependence
of the fitted parameters.

\begin{figure*}[t]\begin{center}
%
\vskip-5.5truecm
\mbox{
\epsfig{figure=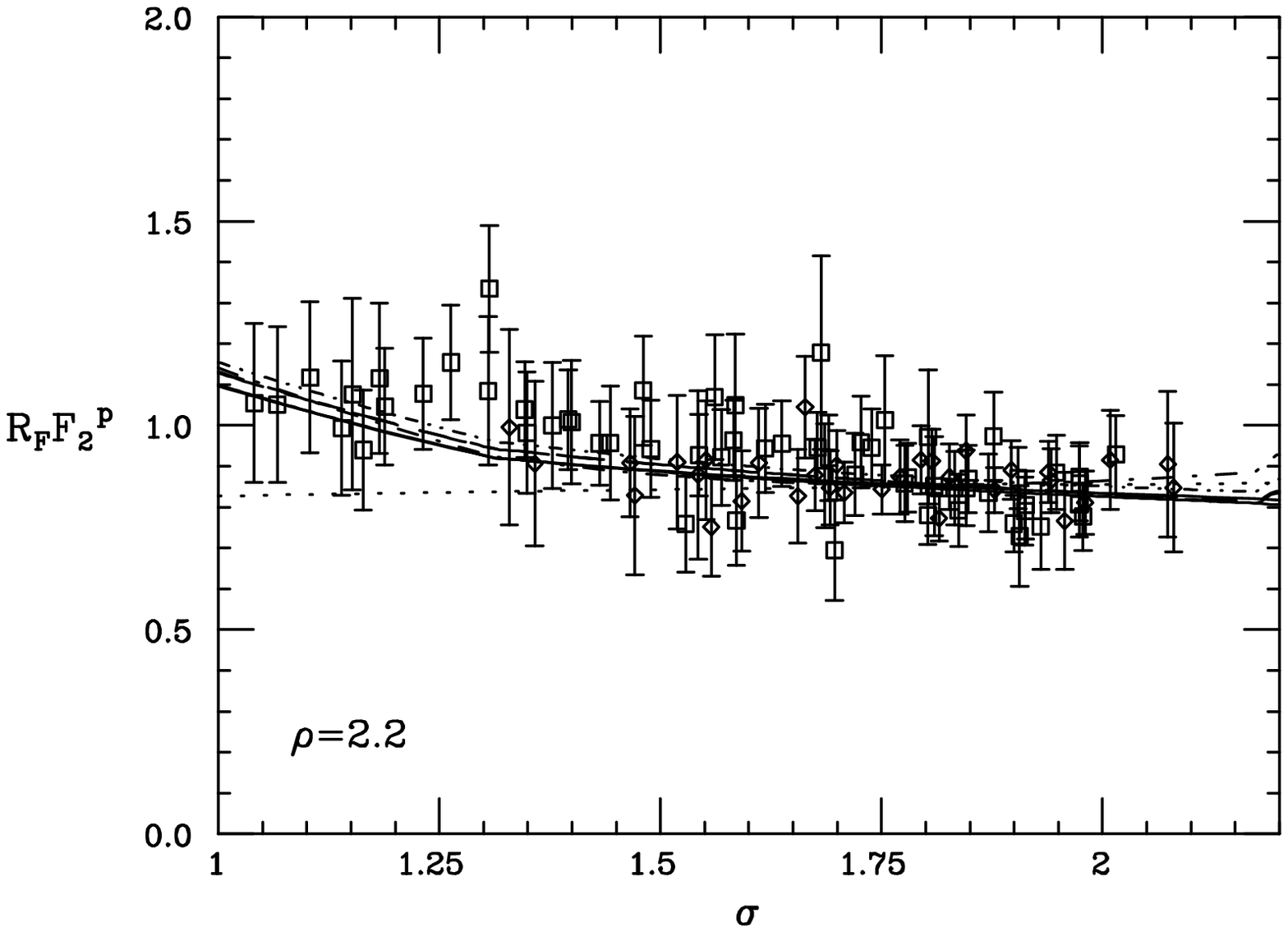,height=16truecm}}
\vskip-6.5truecm
\mbox{
\epsfig{figure=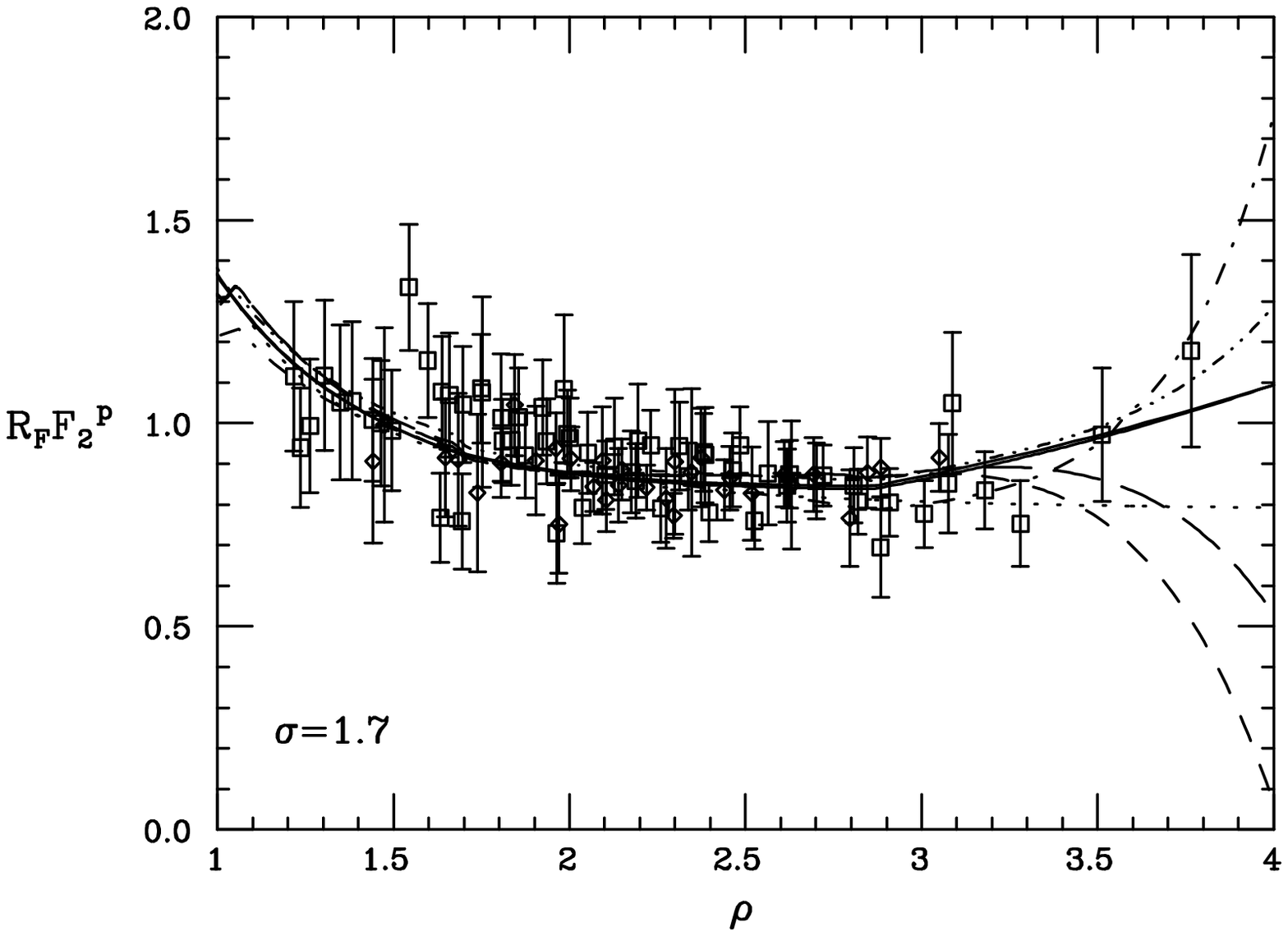,height=16truecm}}
\vskip-4truecm
\caption{Scaling plots corresponding to double scaling
(dotted); two loops (double leading with
$x_0=0$) \MS\ or DIS (solid);  double leading
\MS\ (dot dash), Q$_0$\MS\ (double-dot dash),
Q$_0$DIS (long dashes), DIS (short dashes). The double scaling
curve is from ref.[4],
the other curves correspond to entries
denoted by * in the table.}
\vskip-0.25truecm
\end{center}\end{figure*}

The results of fitting $F_2$ to HERA data \cite{data}
are summarized in the
table and displayed in the figure.\footnote{The
corresponding results of ref.~\cite{alphas} are determined by using a
slightly different treatment of thresholds: here continuity of $F_2$
is imposed (continuity of DIS distributions) whereas there
continuity of the \MS\ parton distributions was required
instead. The slight variation of
the results gives a feeling for the corresponding uncertainty.}
The free parameters are the normalizations of the two data
sets and the small-$x$ exponents of the quark and gluon distributions,
which behave as $x^\lambda$  as $x\to0$;
the resulting $\chi^2$ (for 120 d.f.) is also given.
All fits are performed with $\alpha_s(M_z)=0.120$\cite{alphas};
initial parton distributions are given at $2$~GeV for the $x_0=0$
fits and $3$~GeV for  $x_0=0.1$.\footnote{The starting scale
should also be treated as a
free parameter; it turns out however that a good fit
can be obtained within quite a wide range of values of $Q_0$, the resulting
values of $\lambda$ being decreasing functions of $Q_0$.}

The results can be summarized as follows:\\
a) Whereas the inclusion of two loop corrections improves significantly
the agreement of $F_2$ with the data, going over to the double leading scheme
has very little effect.\\
b) Consequently, the data cannot yet fix the value of $x_0$, however
if $x_0$ is as large as 0.1 they favour $Q_0$ factorization
over the standard one.\\
c) In general both the relative and absolute sizes of
$\lambda_q$, $\lambda_g$ depend strongly
on expansion, fitting, renormalization and factorization schemes.
In particular if $x_0=0$ in \MS\ fitting $\lambda_q\simeq\lambda_g$
(within errors), but in DIS fitting $\lambda_q>\lambda_g$; while if $x_0=0.1$
in \MS\ fitting $\lambda_q >\lambda_g$,
but in DIS fitting $\lambda_q\simeq\lambda_g$.\footnote{This seems to
disagree with
ref.~\cite{mrs} where (on the basis of an \MS\ calculation at two
loops) it is claimed that $\lambda_g$ is significantly smaller than
$\lambda_q$: it also suggests that  some of the assumptions made in the
discussion of the relative size of $\lambda_q$ and $\lambda_g$
in ref.\cite{sdis} are incorrect.}
The exception is that $\lambda_q$ when fitted in DIS is
independent of expansion, renormalization and factorization scheme, since it is
directly related to the small-$x$ behaviour of a physical observable, $F_2$.\\
d)  The scheme dependence of $\lambda_g$ is least severe in
$Q_0$ factorization, where the gluon is generally rather flat.
This provides phenomenological support to the theoretical
expectation\cite{cia} that
the input to perturbative evolution is of more direct physical significance in
this scheme.\footnote{The ``initial Pomeron'' reconstructed from
the best-fit initial gluon distribution according to
ref.\cite{cia} appears then to be soft.}\\
e) Whereas at fixed starting scale
schemes with larger $x_0$ tend to have somewhat smaller values of $\lambda$
(i.e. less singular inputs) the main effect of going over to
the double leading scheme is to reduce the sensitivity to the
starting distribution: double scaling then results
>from a rather wide range of boundary conditions.\\
f) Conversely, there is a very large scheme dependence at large
$\rho$ (i.e. close to the boundary of perturbative evolution) which
may signal a breakdown of leading-twist perturbative calculations there.
This makes a perturbative reconstruction  of the input parton distribution
(and in particular the input gluon) from a measurement
of the evolved structure function very difficult. Which is as it
should be: evolving to  smaller $x$ and/or
lower $Q^2$  leads one eventually
into the intrinsically nonperturbative region.\\
g) Direct measurements of $F_2$ at larger values of $\rho$
 may help to reduce this ambiguity (or at
least postpone it to yet larger values of $\rho$)
by putting constraints  on  $x_0$.
However, if the new data deviate strongly from the two loop
curve this might suggest  a breakdown of leading twist perturbation theory
in this region.

Finally we note that when the physical parameter $\alpha_s$ is also
included in the fit, its value turns out to be largely
insensitive to all of these scheme ambiguities, thereby
allowing a determination of it from small-$x$ structure function
data alone\cite{alphas}.

\bigskip\noindent
{\bf Acknowledgements:} S.F. thanks G.~Altarelli,
S.~Catani, A.~Cooper-Sarkar,
F.~Hautmann, A.~Martin, R.~G.~Roberts and A.~Vogt for interesting
discussions during the conference.

%

\end{document}